\documentclass[journal]{IEEEtran}
\usepackage{amsmath,amsfonts}
\usepackage{algorithmic}
\usepackage{array}
\usepackage[caption=false,font=normalsize,labelfont=sf,textfont=sf]{subfig}
\usepackage{textcomp}
\usepackage{stfloats}
\usepackage{url}
\usepackage{verbatim}
\usepackage{graphicx}
\def\BibTeX{{\rm B\kern-.05em{\sc i\kern-.025em b}\kern-.08em
    T\kern-.1667em\lower.7ex\hbox{E}\kern-.125emX}}
\usepackage{balance}
\usepackage[noadjust]{cite}
\usepackage{capekCommands} % mathematical commands
\usepackage{booktabs}
\usepackage[normalem]{ulem}

\usepackage{xcolor}

\begin{document}
\title{Trade-off Between Optimal Efficiency and Envelope Correlation Coefficient of MIMO Antenna Clusters}
\author{Vojtech Neuman, Miloslav Capek, \IEEEmembership{Senior Member, IEEE}, Lukas Jelinek,\\Anu Lehtovuori, and Ville Viikari, \IEEEmembership{Senior Member, IEEE}
\thanks{Manuscript received \today; revised \today. This work was supported by the Czech Science Foundation under project~\mbox{No.~21-19025M} and by the Czech Technical University in Prague under project~\mbox{SGS22/162/OHK3/3T/13}.}% <-this % stops a space
\thanks{V. Neuman, M. Capek and L. Jelinek are with the Czech Technical University in Prague, Prague, Czech Republic (e-mails: \{vojtech.neuman; miloslav.capek; lukas.jelinek\}@fel.cvut.cz).}% <-this % stops a space
\thanks{A. Lehtovuori and V. Viikari are with the Aalto University, Espoo, Finland (e-mails: \{anu.lehtovuori; ville.viikari\}@aalto.fi).}% <-this % stops a space
\thanks{Color versions of one or more of the figures in this paper are available online at http://ieeexplore.ieee.org}
}

\markboth{}%
{How to Use the IEEEtran \LaTeX \ Templates}

\maketitle

\begin{abstract}
The presented paper proposes a theory for the optimization of multiple-input-multiple-output antenna performance by assessing feeding coefficients. Antenna clusters with multiple feeding ports are utilized, which brings additional degrees of freedom and improves the performance. This work considers fixed shape and matching networks. The method is based on quadratic programming and maximizes total efficiency constrained by channel correlation and channel power distribution. The formulation provided in the paper enables establishing trade-offs between all mentioned metrics. Evaluating the performance in this manner provides comprehensive information about the chosen geometry and port placement. Selected examples demonstrate how efficiency and channel correlation can be both in agreement but also in significant conflict. The effect of frequency dispersion on feeding is also investigated.
\end{abstract}


%%%%%%%%%%%%%%%%%%%%%%%%%%%%%%%%%%%%%%%%%%%%%%%%%%%%%%%%%%
%                           #   
%                          ##   
%                         # #   
%                           #   
%                           #   
%                           #   
%                         ##### 
%%%%%%%%%%%%%%%%%%%%%%%%%%%%%%%%%%%%%%%%%%%%%%%%%%%%%%%%%%
\section{Introduction}
\label{sec:Introduction}

\IEEEPARstart{M}{ultiple-input-multiple-output} (MIMO) technology has proven its utility in wireless communication in increasing data throughput and coverage while mitigating signal fading due to multipath propagation~\cite{book:PaulrajIntroSpaceTimeWireComm}. All contemporary wireless standards utilize MIMO~\cite{web:3GPP}, and this trend is expected to continue~\cite{art:Giordani2020Toward6GNets, art:Akyildiz20206GBeyond}. Of high priority is the implementation of MIMO technology for mobile devices at sub-6\,GHz frequency bands~\cite{web:3GPP}. Though narrow bands at this frequency range cannot offer data rates as high as millimeter-wave frequencies, they still have an advantage in robustness against environmental influences and free space loss~\cite{book:GhasemiPropagEngWirelessComm, art:Shafi2018MicroVsMiliWavePropagation}. The small electric size of user devices in these frequency bands nevertheless significantly degrades their efficiency and channel correlation, two key parameters affecting MIMO system performance~\cite{book:PaulrajIntroSpaceTimeWireComm,art:Vaughan1987AntennaDiversity}.

Efficiency and channel correlation are closely associated with mutual coupling between antennas. As the mutual coupling reduction improves both mentioned quantities, several approaches were introduced to suppress it~\cite{art:Kumar2020FifthGenerationAntennas, art:Nadeem2019StudyOnMutCoupling}. The ground plane is usually the main radiator at low frequencies. Introducing various slits and defects into it can improve isolation between ports~\cite{art:Aloi2012A4ShapedMultiStandardCompactMIMO, art:Ayatollahi2012CompactHighIsolationWideBandAntenna, art:Chiu2007ReductionMutualCoup, art:Ding2007FourElemAntSyst, art:Dong2017DecoupledMultibandDualAntenna, art:Sharawi2011An800MHzCompactMIMO, art:Zhang2008CompactMIMOAnt}. Another method is based on the application of decoupling networks between signal generators and antenna ports~\cite{art:Andersen1976DecoupDescatterNetworks, art:Chen2008DecoupTechniquePortIsol, art:Wu2013StubLoadReacDecoupNetwork}. A similar principle occurs with neutralization lines, where the mutual coupling is reduced by connecting transmission line of appropriate length between elements~\cite{art:Dumanli2011SlotAntennaArrayLowMutualCoup, art:Farsi2012MutualCouplingReductionPlanarAntennas, art:Su2012PrintedMIMOAntennaSystem, art:Wang2014UltraCompactThreePort}. Diversity between antennas can also be achieved by using different far-field polarization of each channel~\cite{art:Ren2019ACompactBuildingBlockEightAntennaMIMO, art:Wang2009PatternPolarDiversityAntenna, art:Wi2013MultibandHandsetAntenna, art:Lee1972PolarDiversitySystRadio}. Alternative to previous methods is the utilization of characteristic modes~\cite{art:Harrington1971TheoryCharModes}. Information contained in modal analysis~\cite{art:CabedoFabres2007TheoryCharModesRevisited} provides a powerful tool for port placement for the excitation~\cite{art:Li2018APatternReconfigurableMIMO, art:Liu2019EightPortMIMOArrayCharModes, art:Manteuffel2016CompactMultiAntenna, inp:Martens2012AntBasedSelectCharModes} of orthogonal far-fields or can be used for driving shape modifications~\cite{art:Adams2022AntennaElementDesignUsingCharModes, art:Yang2016SystShapeOptCharModes, art:Martens2014SystDesignMethod} to obtain the desired performance. Further utilization of modal properties and symmetries lead to establishing some fundamental limits on MIMO performance~\cite{art:Masek2021ExctOrthoRadStates, art:Peitzmeier2019UpperBoundUncorrPort}. The listed methods have been proven to be capable of enhancing MIMO performance. However, their utilization is accompanied by several challenges. Structure modifications and reactive networks both suffer from demanding design processes and assessing optimal suppressing structure might be almost impossible. Another problem lies in narrow bandwidths and frequency reconfigurability. The characteristic modes additionally encounter issues with selective excitation~\cite{inp:Martens2011IndCapExctCharModes}.

To deal with all the mentioned difficulties, this work establishes its proposed solution on the utilization of so-called antenna clusters technology~\cite{art:Hannula2017FreqReconfigMultibandAntenna}. One antenna cluster groups multiple radiating elements, each having its active feed. In addition to structural modifications and matching circuitry, the performance of the antenna cluster can further be improved by employing feeding optimization. In this paradigm, mutual coupling between individual ports within one cluster does not have to be suppressed and is beneficial for operation instead~\cite{inp:Hannula2019BenefInteractCoupling}. Using multiple diverse elements within the antenna cluster enables frequency reconfigurability~\cite{art:Hannula2017ConceptFreqReconfigAntenna}. The feeding circuitry is replaced by advanced solutions represented by distributed transmitter~\cite{art:Saleem2020IntegratedRFTransmitter}. Though the antenna cluster has multiple ports, it constitutes only one communication channel. Using multiple antenna clusters in the design region enables MIMO operation~\cite{art:Hannula2019TunEightAntenna}. Mutual coupling between two antenna clusters still causes problems and has to be suppressed either by modifications in the radiating system or by the utilization of additional degrees of freedom in multiple ports. Nevertheless, no previous work considered the additional ports for channel correlation mitigation, and any method for setting feeding coefficients was not proposed.

This paper introduces a theoretical framework for feeding optimization of total efficiency and channel correlation, constrained by a user-defined allocation of power radiated by channels. The former problem for maximization of total efficiency~\cite{art:Capek2020FindOptTotalRefCoef, art:Hannula2017FreqReconfigMultibandAntenna} is supplemented by additional constraints capable of dealing with the unwanted components in radiated power. The proposed theory utilizes matrix description and shows how radiated power can be decomposed into individual parts. Described tools allow comprehensive evaluation of trade-offs between desired parameters and decisions about further design processes. This work assumes the radiator's shape and matching circuitry to be fixed, and only the feeding coefficients are manipulated to assess fundamental limits on the excitation of the given antenna design. The definition of problems enables direct calculation of the optimal feeding weights. All listed methods for MIMO design are compatible with the antenna clusters technology and can be used in arbitrary combinations.

The organization of this paper is as follows. Section~\ref{sec:MathematicalBackground} introduces the mathematical tools used. The modified optimization procedure is described in Sec.~\ref{sec:OptimizationProbConst}, and its application is shown in Sec.~\ref{sec:Example} and Sec.~\ref{sec:ExampleII}. The paper concludes in Sec~\ref{sec:Conclusion}. Appendices~\ref{sec:MatrixOperators} and~\ref{sec:FarFieldCorr} reviews the theoretical background, and Appendices~\ref{sec:SystemAnalysis} and~\ref{sec:RadPowerRatios} deal with further aspects of the developed theory.

%%%%%%%%%%%%%%%%%%%%%%%%%%%%%%%%%%%%%%%%%%%%%%%%%%%%%%%%%%
%                          #####  
%                         #     # 
%                               # 
%                          #####  
%                         #       
%                         #       
%                         ####### 
%%%%%%%%%%%%%%%%%%%%%%%%%%%%%%%%%%%%%%%%%%%%%%%%%%%%%%%%%%
\section{Antenna Cluster Performance Evaluation}
Considering the antenna's role in the communication chain, the channel capacity is influenced by the three following parameters~\cite{art:Kildal2004CorrelationAndCapacity, art:Tian2011MultiplexEfficiency, art:Vaughan1987AntennaDiversity}: total efficiency, envelope correlation coefficient, and power allocation among individual channels. Additionally, the complete assessment of the channel capacity requires selecting a channel model and a decision about the used transmitting and receiving antennas. The described process is out of the scope of this work, and only the three antenna parameters mentioned above will be addressed further.

Numerous works deal with total efficiency evaluated for each channel individually---the channel's efficiency is assessed by feeding that channel and terminating the rest of the channels into the standard load~\cite{art:Yun2012MultiElementAntennaEfficiency}. This approach allows straightforward verification by measurement. The presented work searches for the maximum attainable performance. Therefore, the former method is replaced by evaluating total efficiency when all ports are synchronously driven with the signals. This second approach includes losses caused by coupling between individual channels and fully considers the effects of mutual coupling on radiated power~\cite{art:Wallace2004MutCouplingMIMO}.

\begin{figure}[t]
\centering
\includegraphics[scale=0.95]{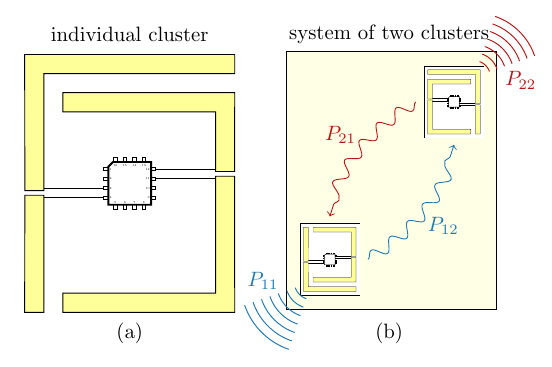}
\caption{Illustration of an antenna cluster. The indicated integrated circuit represents a distributed transceiver. (a) Multi-port antenna forming an antenna cluster (two feeding ports, two radiating elements). (b) The MIMO system consisting of two antenna clusters. The wavy curves and arrows depict self and mutual radiated powers.}\label{pic:AntennaClusterPic}
\end{figure}

The total efficiency and envelope correlation coefficients can be optimized by shape modifications or the usage of reactive loading. The power allocation also depends on these two methods. However, it can be scaled arbitrarily by feeding networks. Besides the mentioned approaches, it is further possible to employ auxiliary ports serving as additional degrees of freedom. In this work, the whole radiating system consists of $N$ ports divided into $M$ groups, where $m$-th group has $N_m$ ports, where $M$ also defines the intended order of MIMO functionality. A group of ports is called an antenna cluster, see Fig.~\ref{pic:AntennaClusterPic}(a), and this nomenclature is held throughout the paper. The MIMO system established by multiple antenna clusters is illustrated in~Fig.~\ref{pic:AntennaClusterPic}(b). The dependence of antenna clusters on applied feeding coefficients is exploited below for optimization purposes.

The operation of antenna clusters is supported by distributed transmitters~\cite{art:Saleem2020IntegratedRFTransmitter}, which sets desired feeding coefficients. The realistic functionality of this component is not taken into account in this work and it is assumed that feeding coefficients can be set arbitrarily. Depending on the designer's choice, the feeding coefficients might be port voltages, port currents, or incident power waves. Subsequent work is derived on the usage of port voltages. The alternative with incident power waves is outlined in Appendix~\ref{sec:PowerWavesForm}.

%%%%%%%%%%%%%%%%%%%%%%%%%%%%%%%%%%%%%%%%%%%%%%%%%%%%%%%%%%
%                          #####  
%                         #     # 
%                               # 
%                          #####  
%                               # 
%                         #     # 
%                          #####  
%%%%%%%%%%%%%%%%%%%%%%%%%%%%%%%%%%%%%%%%%%%%%%%%%%%%%%%%%%
\section{Radiation of Antenna Clusters}
\label{sec:MathematicalBackground}
The radiating system is fully described with its port radiation matrix $\pGmatFS\in\mathbb{C}^{N\times N}$, see Appendix~\ref{sec:MatrixOperators}, which, together with port voltages, calculate radiated power
\begin{equation}
\frac{1}{2}\pVvecH\pGmatFS\pVvec = \Prad,
\end{equation}
where $\Prad$ denotes time-averaged radiated power, and~$\pVvec\in\mathbb{C}^{N\times 1}$ aggregates feeding voltages connected to the antenna ports. Considering the far-field correlation, see Appendix~\ref{sec:FarFieldCorr}, the port radiation matrix $\pGmatFS$ is the operator of primary interest\footnote{Although the lossless antennas are used for simplicity, lossy cases can be included without any major modification of the theory, see Appendix~\ref{sec:MatrixOperators}.}. Individual elements of radiation matrix~$\pGmatFS$ are grouped into blocks as
\begin{equation}
\pGmatFS = 
\begin{bmatrix}
\pGmatFSind{1}{1} & \hdots & \pGmatFSind{1}{M} \\
\vdots & \ddots & \vdots \\
\pGmatFSind{M}{1} & \hdots & \pGmatFSind{M}{M}
\end{bmatrix},\label{eq:g0mat}
\end{equation}
with \mbox{$\pGmatFSind{m}{m} \in \mathbb{C}^{N_m\times N_m}$} describing the interaction within the $m$-th cluster and \mbox{$\pGmatFSind{m}{n}\in \mathbb{C}^{N_m\times N_n}$} considering interactions between the $m$-th and $n$-th cluster. The radiation matrix is hermitian
\begin{equation}
\pGmatFSind{n}{m} = \pGmatFSind{m}{n}^\herm.\label{eq:GmatHermit}
\end{equation}

The feeding of an antenna is also partitioned into blocks, each representing a feeding vector of a particular cluster
\begin{equation}
\pVvec = \begin{bmatrix} \pVvec_1^\trans & \hdots & \pVvec_M^\trans \end{bmatrix}^\trans,
\end{equation}
where $\pVvec \in \mathbb{C}^{N\times 1}$ contains vectors $\pVvec_m \in \mathbb{C}^{N_m\times 1}$ corresponding to the $m$-th cluster feeding.

The radiated power consists of terms 
\begin{equation}
P_{mn} = \frac{1}{2}\pVvecH_m\pGmatFSind{m}{n}\pVvec_n,\label{eq:RadPowerTerms}
\end{equation}
where $m=n$ represents self terms, and $m\neq n$ represents interaction terms. Summing all contributions gives the total radiated power
\begin{equation}
\Prad = \frac{1}{2}\sum_{m} \pVvecH_m\pGmatFSind{m}{m}\pVvec_m + \frac{1}{2}\sum_{m}\sum_{n\neq m} \pVvecH_m\pGmatFSind{m}{n}\pVvec_n. \label{eq:TotalRadiatedPower}
\end{equation}
Notice that only total radiated power~$\Prad$ in~\eqref{eq:TotalRadiatedPower} has a clear physical meaning. The separation into generally complex-valued terms~\eqref{eq:RadPowerTerms} is only used to exploit their mathematical properties for the optimization procedure discussed in Sec.~\ref{sec:OptimizationProbConst}. It is nevertheless worth mentioning that power terms $P_{mn}$ are directly related to the far-field correlation coefficient, see~\eqref{eq:FarFieldCorrelationVolt} in Appendix~\ref{sec:FarFieldCorr}.

Dividing relation~\eqref{eq:TotalRadiatedPower} by total radiated power $\Prad$ leads to power ratios~$\alpha_{mn}$
\begin{multline}
1 = \sum_{m}\alpha_{mm} + \sum_{m}\sum_{n\neq m}\alpha_{mn} = \\ \sum_{m}\alpha_{mm}  + \sum_{m}\sum_{n > m}\beta_{mn}, \label{eq:PowerCoefSummation}
\end{multline}
where $\alpha_{mn} = P_{mn}/\Prad$ and $\beta_{mn} = \alpha_{mn} + \alpha_{nm}$. Power ratios $\alpha_{mn}$ with $m\neq n$ have an imaginary part, for which $\gamma_{mn}$ is defined as
\begin{equation}
\gamma_{mn} = \I\left(\alpha_{mn} - \alpha_{nm}\right),
\end{equation}
where $\I^2 = -1$ represents imaginary unit. The power ratios are used in the optimization to control the composition of the radiated power. The self-term power ratios~$\alpha_{mm}$ control useful radiated power, and $\beta_{mn},\gamma_{mn}$ are used to limit the mutual radiated power. More details are given in Appendices~\ref{sec:SystemAnalysis} and~\ref{sec:RadPowerRatios}.

%%%%%%%%%%%%%%%%%%%%%%%%%%%%%%%%%%%%%%%%%%%%%%%%%%%%%%%%%%
%                         #       
%                         #    #  
%                         #    #  
%                         #    #  
%                         ####### 
%                              #  
%                              #  
%%%%%%%%%%%%%%%%%%%%%%%%%%%%%%%%%%%%%%%%%%%%%%%%%%%%%%%%%%
\section{Optimization Problem and Constraints}
\label{sec:OptimizationProbConst}

The trade-off between far-field correlation, total efficiency, and power allocation is formulated in this section, starting with simpler sub-problems.

The optimization problem for maximum total efficiency~$\toteff$ reads 
\begin{equation}
\begin{split}
\underset{\pVvec}{\T{maximize}} \quad & \Prad \\
\T{subject\, to} \quad & \Pin = 1, \\
\end{split}
\label{eq:OptTaskReference}
\end{equation}
and it can be solved as a generalized eigenvalue problem~\cite{art:Capek2020FindOptTotalRefCoef}. This formulation delimits the highest total efficiency achieved with a given set of ports. To constrain optimization, the power delivered to the antennas is fixed see Appendix~\ref{sec:MatrixOperators} with further explanation. The problem~\eqref{eq:OptTaskReference} is unable to distinguish individual channels. Therefore, a modification in the form of additional constraints is required.

Observing the formula for the far-field correlation~\eqref{eq:FarFieldCorrelationVolt}, it is clear that the term in its numerator should be minimized. The terms of form~$P_{mn}$ with $m\neq n$ are generally complex and unsuitable for convex solvers. Thus\RR{,} real mutual radiated power $\PradMutReal{m}{n}$ is introduced and defines a constraint on mutual radiated power
\begin{equation}
\PradMutReal{m}{n} = 2\RE\left\{P_{mn}\right\} = \beta_{mn}\Prad.\label{eq:MutRealPowerConstr}
\end{equation}
where $\beta_{mn}$ is the power ratio from~\eqref{eq:PowerCoefSummation}, the value set by the user during optimization.

Constraint~\eqref{eq:MutRealPowerConstr} deals only with the real part of the correlation coefficient~\eqref{eq:FarFieldCorrelationVolt}. The imaginary part also has to be suppressed\footnote{Real part of $P_{mn}$ term expresses the time-averaged power flowing from $m$-th cluster to $n$-th cluster. The imaginary part does not have a direct physical explanation but is an important part of the correlation coefficient.}. A constraint on imaginary mutual power $\PradMutImag{m}{n}$ reads
\begin{equation}
\PradMutImag{m}{n} = 2\IM\left\{P_{mn}\right\} = \gamma_{mn}\Prad\label{eq:MutImagPowerConstr}
\end{equation}
where $\gamma_{mn}$ is used in a similar manner as $\beta_{mn}$.

Optimization problem~\eqref{eq:OptTaskReference} constrained solely by relations~\eqref{eq:MutRealPowerConstr} and~\eqref{eq:MutImagPowerConstr} would end in a trivial solution, where some $\pVvec_m$ would be a zero vector, which would mean that corresponding $\Prad_{mm}$ vanishes. Such a scenario is unacceptable since it disables one of the clusters (and with it the corresponding communication channel). Hence, another set of constraints must be defined. To that point, self-radiated powers $\PradSelf{m}$ are constrained via
\begin{equation}
\PradSelf{m} = \alpha_{mm}\Prad,\label{eq:SelfPowerConstr}
\end{equation}
stating that the channel radiates a user-defined ratio~$\alpha_{mm}$ of total radiated power.

Considering the constraints defined above, the modified optimization problem reads
\begin{equation}
\begin{split}
\underset{\pVvec}{\T{maximize}} \quad & \Prad \\
\T{subject\, to} \quad & \Pin = 1, \\
\quad & \PradSelf{m} = \alpha_{mm}\Prad,\,\forall m\\
\quad & \PradMutReal{m}{n} = \beta_{mn}\Prad,\,\forall m,n \wedge m\neq n,\\
\quad & \PradMutImag{m}{n} = \gamma_{mn}\Prad,\,\forall m,n \wedge m\neq n,
\end{split}
\label{eq:OptTaskConstrained}
\end{equation}
This definition fully controls radiated power optimization and far-field correlation while maximizing total efficiency.

Before showing the described theory in an example, it is worth pointing out the following remarks. The system with $M$ clusters needs, in total, $M^2$ constraints\footnote{The given number respects the dependence of power components.}. This number becomes considerably high even for a small number of antenna clusters. In practice, however, some constraints might be omitted. For example, the channel correlation between particular clusters can be already sufficiently low due to a geometrical arrangement.

Although the first choice of the power ratios in problem~\eqref{eq:OptTaskConstrained} might be $\alpha_{mm} = 1 / M$ and $\beta_{mn},\gamma_{mn} \rightarrow 0$, it is beneficial to perform a sweep over these values. The sweep not only provides an understanding of how radiated power is distributed among the channels, but it might also offer important relaxations that would be compensated by gains in overall performance.

Throughout this paper, the solution of problem~\eqref{eq:OptTaskConstrained} is approached using a dual formulation~\cite{book:BoydConvOpt} and quadratically constrained quadratic program solvers contained within the fundamental bounds package~\cite{misc:Liska2021FundamentalBounds}. When a duality gap~\cite{book:BoydConvOpt} is present, the interior-point method~\cite{book:NocedalNumerOptim} applied directly to the primal problem is utilized instead. In these cases, the localization of a global optimum is not assured. A comparison with the Monte Carlo method suggests that these extremal points might be globally optimal.

The problem~\eqref{eq:OptTaskConstrained} is general and can be used for arbitrary complex scenarios, resolved by the arbitrary full-wave electromagnetic solver since the matrix describing radiation properties can always be obtained from far-field integration~\cite{art:Stein1962OnCrossCoupling}. Definition~\eqref{eq:OptTaskConstrained} further allows the integration of knowledge about the feeding circuitry with the method introduced in~\cite{art:Capek2023TransRadEfficiency,art:Broyde2022RadTransEfficiencies} by replacing the incident power~\cite{book:PozarMicroEng} with available power~\cite{art:Desoer1973MaxPowerTransfer}. Appendix~\ref{sec:PowerWavesForm} shows the modification of problem~\eqref{eq:OptTaskConstrained} when the optimization variables and matrices are represented by power waves.

%%%%%%%%%%%%%%%%%%%%%%%%%%%%%%%%%%%%%%%%%%%%%%%%%%%%%%%%%%
%                         ####### 
%                         #       
%                         #       
%                         ######  
%                               # 
%                         #     # 
%                          #####  
%%%%%%%%%%%%%%%%%%%%%%%%%%%%%%%%%%%%%%%%%%%%%%%%%%%%%%%%%%
\section{Example: Parallel Dipoles}
\label{sec:Example}

\begin{figure}[t]
\centering
\includegraphics[scale=0.95]{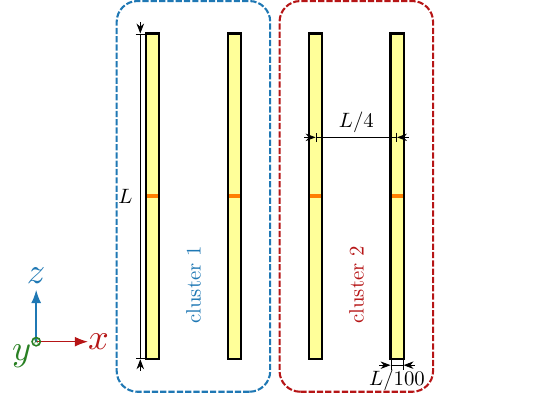}
\caption{Illustration of the considered MIMO system. The dimensions are not shown in the correct proportions. The blue and red curves highlight the first and second clusters. All dipoles are fed with delta gaps in places highlighted with orange lines.}\label{pic:Parallel4Dipole}
\end{figure}

To demonstrate the most salient features of the proposed optimization, four thin parallel dipoles made of a perfect electric conductor are used, see Fig.~\ref{pic:Parallel4Dipole}. The length of each strip is equal to $L=0.916\,c_0/\left(2f_0\right)$, where $f_0 = 750\,$MHz. Spacing between dipoles is equal to $L/4$ and is purposely chosen to highlight problems with an envelope correlation coefficient between considered antenna clusters. All ports are connected to the $50\,\Omega$ transmission line, and no additional components are used for matching. Figure~\ref{pic:Parallel4Dipole} also shows how antenna elements are grouped into two clusters. The in-house method of moments solver~\cite{web:AToM} is employed for the system analysis. 

The optimization problem reads
\begin{equation}
\begin{split}
\underset{\pVvec}{\T{maximize}} \quad & \Prad \\
\T{subject\, to} \quad & \Pin = 1, \\
\quad & \PradSelf{1} = \alpha_{11}\Prad, \\
\quad & \PradMutReal{1}{2} = \beta_{12}\Prad, \\
\quad & \PradMutImag{1}{2} = \gamma_{12}\Prad,
\end{split}
\label{eq:OptTask4Dipole}
\end{equation}
where $\alpha_{11},\beta_{12}$, and $\gamma_{12}$ are swept. In practical scenarios, a given ratio~$\PradRatio$ between $\PradSelf{1}$ and $\PradSelf{2}$ is typically of interest. Utilizing the dependence~\eqref{eq:PowerCoefSummation} and defining a power ratio $\PradRatio$ as
\begin{equation}
\PradRatio = \frac{P_{22}}{P_{11}} = \frac{\alpha_{22}}{\alpha_{11}}
\end{equation}
allows fixing the power ratio~$\PradRatio$ by using
\begin{equation}
\alpha_{11} = \frac{1 - \beta_{12}}{1 + \PradRatio},
\end{equation}
thereby reducing the sweep to only two dimensions. The rest of this section is dedicated to solving problem~\eqref{eq:OptTask4Dipole} on different frequencies with various power ratio settings to show different features of the developed theory. 

\begin{figure}[t]
\centering
\includegraphics[width=0.48\textwidth]{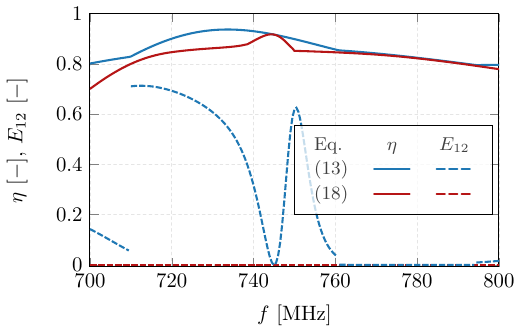}
\caption{The frequency sweep of total efficiency~$\toteff$ and envelope correlation coefficient~$E_{12}$ for $\PradRatio=1$, $\beta_{12} = 0$ and $\gamma_{12} = 0$. The blue color represents the former optimization problem~\eqref{eq:OptTaskReference}, and the red is its modified version~\eqref{eq:OptTask4Dipole}. The jump in the blue dashed curve is caused by the change of feeding eigenvector~$\pVvec$ to a different mode.}\label{pic:Parallel4DipoleTotEffAndECC}
\end{figure}

The results of the two optimization problems are compared: the maximization of total efficiency~$\eta$ in problem~\eqref{eq:OptTaskReference} and the simultaneous minimization of the envelope correlation coefficient $E_{12}$ in problem~\eqref{eq:OptTask4Dipole}. In the latter case, the optimization is solved using~$\PradRatio=1$, $\beta_{12} = 0$ and $\gamma_{12} = 0$, \ie{}, enforcing the zero correlation coefficient and enforcing equal power distribution among the channels. Considering the properties of constrained optimization~\cite{book:BoydConvOpt,book:NocedalNumerOptim}, the expected result is a lower value of total efficiency for constrained problem~\eqref{eq:OptTask4Dipole} as compared to the former problem~\eqref{eq:OptTaskReference}. Figure~\ref{pic:Parallel4DipoleTotEffAndECC} shows that optimization of total efficiency leads to a relatively large envelope correlation coefficient ($E_{12} > 0.5$) in most of the studied frequencies. When the constraint on $E_{12}$ is added, total efficiency is slightly decreased, but the envelope correlation coefficient is kept at zero guaranteeing acceptable MIMO performance.

\begin{figure}[t]
\centering
\includegraphics[width=0.48\textwidth]{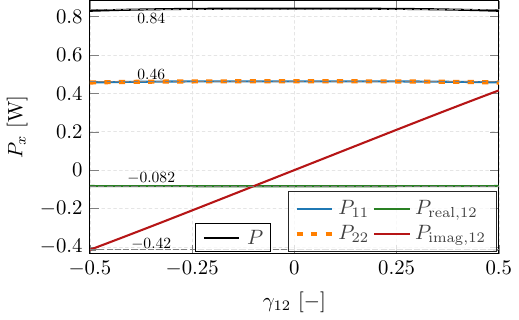}
\caption{The sweep of power ratio~$\gamma_{12}$ at $f = 721\,$MHz with~$\beta_{12} = -0.1$,  $\alpha_{11} = 0.55$ and $\alpha_{22} = 0.55$. The input power reads $\Pin = 1$\,W. The black curve denotes total radiated power~$P$. The rest of the curves depict its decomposition into individual parts~\eqref{eq:TotalRadiatedPower}.}\label{pic:Parallel4DipoleGammaSweep}
\end{figure}

Various power ratio settings are further studied in Fig.~\ref{pic:Parallel4DipoleGammaSweep} which demonstrates the situation in which problem~\eqref{eq:OptTask4Dipole} is solved at $f=721\,$MHz with~$\PradRatio = 1$, $\beta_{12} = -0.1$ and $\gamma_{12}\in\left[-0.5,0.5\right]$. It shows how different power components~\eqref{eq:TotalRadiatedPower} are related to total radiated power $P$ through power ratios. The curves for $P_{11}$ and $P_{22}$ overlay each other as the power ratio reads~$\PradRatio = 1$.

\begin{figure}[t]
\centering
\includegraphics[width=0.48\textwidth]{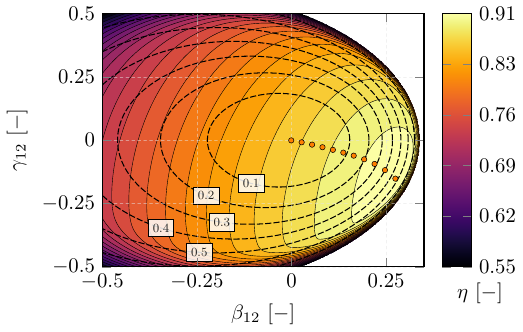}
\caption{The power ratio sweep $\beta_{12}\in\left[-0.5,0.5\right]$ and $\gamma_{12}\in\left[-0.5,0.5\right]$ for power ratio $\PradRatio=10$. The solid-line contour plot depicts total efficiency. Dashed ellipses are regions of constant envelope correlation~$E_{12}$. The orange dots refer to Pareto-optimal solutions Fig.~\ref{pic:Parallel4DipoleTradeOffTotEffECC}.}\label{pic:Parallel4DipoleTotEffCircleECC}
\end{figure}

Evaluating the trade-off between total efficiency~$\toteff$ and envelope correlation coefficient~$E_{12}$, the power ratio sweep is extended to~$\beta_{12},\gamma_{12}\in\left[-0.5,0.5\right]$, see Fig.~\ref{pic:Parallel4DipoleTotEffCircleECC}, and problem~\eqref{eq:OptTask4Dipole} is solved at~$f=735\,$MHz with~$\PradRatio=10$. Total efficiency is depicted as a contour plot and envelope correlation is shown with ellipses of constant $E_{12}$, see Appendix~\ref{sec:SystemAnalysis}. The figure also shows that not all combinations of power ratios lead to a feasible solution. The unfeasible solutions are bounded by a maximal envelope correlation ellipse. Optimal total efficiency and the optimal envelope correlation coefficient lie at different spots in the $\beta_{12}\times\gamma_{12}$ space, showing that they are conflicting parameters. 

\begin{figure}[t]
\centering
\includegraphics[width=0.48\textwidth]{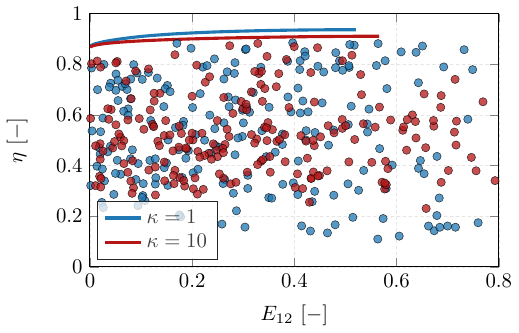}
\caption{The trade-off between the envelope correlation coefficient and total efficiency for two different ratios~$\PradRatio$ between self powers. Pareto optimal curves are highlighted. Dots represent randomly generated solutions.}\label{pic:Parallel4DipoleTradeOffTotEffECC}
\end{figure}

The trade-off between total efficiency~$\toteff$ and envelope correlation coefficient~$E_{12}$ at frequency $f=735\,$MHz is shown in Fig.~\ref{pic:Parallel4DipoleTradeOffTotEffECC}. The blue and red continuous curves are Pareto optimal solutions. Validation of the proposed framework is supported using the Monte Carlo method to generate random feeding vectors. All voltage vectors $\pVvec$ are normalized to satisfy constraints on input power and the power ratio~$\PradRatio$. The blue and red dots represent those solutions, and all lie under the plotted Pareto frontiers. In this specific case, there is only a slight difference in performance between different power ratios $\PradRatio$. Also, the drop in efficiency for lower envelope correlation coefficients is minor.

\begin{figure}[t]
\centering
\includegraphics[scale=1]{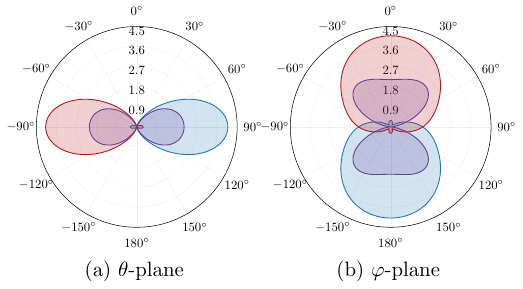}
\caption{Total and partial radiation patterns resulting from the solution to~\eqref{eq:OptTask4Dipole}  for $\PradRatio = 1$, $\beta = -0.045$ and $\gamma=0$ at central frequency~$f_0$. The red and blue colors represent each channel. (a) cut in $\varphi=0$. (b) cut in $\theta=\pi/2$.}\label{pic:Parallel4DipoleFarField}
\end{figure}

The radiation patterns for~$\PradRatio = 1$, $\beta = -0.045$ and $\gamma=0$, see Fig.~\ref{pic:Parallel4DipoleTotEffAndECC}, at frequency~$750\,$MHz are depicted in Fig.~\ref{pic:Parallel4DipoleFarField}. The pattern of the whole system corresponding to voltage vector~$\pVvec = \begin{bmatrix} \pVvec_1^\trans & \pVvec_2^\trans  \end{bmatrix}^\trans$ solving problem~\eqref{eq:OptTask4Dipole} is depicted in purple. This radiation pattern is composed of individual patterns of each cluster (blue and red), which correspond to voltage vectors~$\pVvec = \begin{bmatrix} \pVvec_1^\trans & \M{0}^\trans  \end{bmatrix}^\trans$ and~$\pVvec = \begin{bmatrix} \M{0}^\trans & \pVvec_2^\trans  \end{bmatrix}^\trans$, respectively, and resemble cardioids with the opposite orientation of their main lobes, a sound solution for high efficiency and low far-field correlation.

%%%%%%%%%%%%%%%%%%%%%%%%%%%%%%%%%%%%%%%%%%%%%%%%%%%%%%%%%%
%                              # 
%                            #       
%                           #       
%                          #####  
%                         #     # 
%                         #     # 
%                          #####  
%%%%%%%%%%%%%%%%%%%%%%%%%%%%%%%%%%%%%%%%%%%%%%%%%%%%%%%%%%
\section{Example: Mobile Terminal}
\label{sec:ExampleII}

\begin{figure}[t]
\centering
\includegraphics[scale=1]{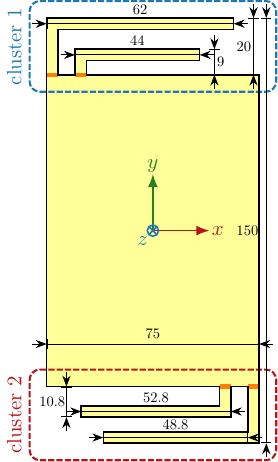}
\caption{Simplified model of a mobile terminal. The blue and red curves highlight the first and second clusters. Monopole antennas are fed with delta gaps in places highlighted with orange lines. The dimensions are in millimeters.}\label{pic:Cellphone4Port}
\end{figure}

The second studied structure resembles a mobile terminal introducing a more realistic antenna example with size restricted to common present-day devices, see Fig.~\ref{pic:Cellphone4Port}. The antenna system is considered for operations in the vicinity of~$900$\,MHz. All ports are connected to~$50\,\Omega$ transmission lines and no additional components are used for matching. The presented system is purposely not symmetric. The four antenna elements are grouped into two clusters as indicated in Fig.~\ref{pic:Cellphone4Port}.

Enforcing the vanishing envelope correlation coefficient is, generally, not the best strategy for optimization due to its price in total efficiency. Instead, it is beneficial to add an additional non-equality constraint that sets a tolerance for envelope correlation coefficient~$E_{12}$ under a given threshold
\begin{equation}
\begin{split}
\underset{\pVvec}{\T{maximize}} \quad & \Prad \\
\T{subject\, to} \quad & \Pin = 1, \\
\quad & \PradSelf{1} = \alpha_{11}\Prad, \\
\quad & \PradMutReal{1}{2} = \beta_{12}\Prad, \\
\quad & \PradMutImag{1}{2} = \gamma_{12}\Prad, \\
\quad & E_{12} \leq E_\T{max}, \\
\end{split}
\label{eq:OptTask4Cellphone}
\end{equation}
where $E_\T{max}$ is the maximal allowed envelope correlation coefficient. 

\begin{figure}[t]
\centering
\includegraphics[width=0.48\textwidth]{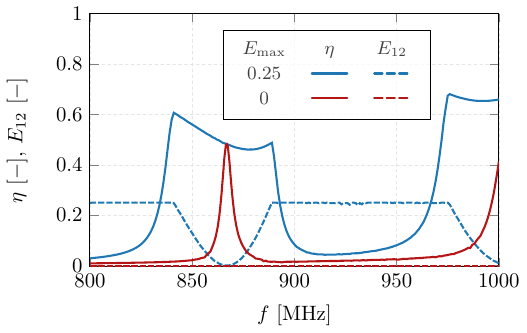}
\caption{The frequency sweep of total efficiency~$\toteff$ and the envelope correlation coefficient~$E_{12}$ for $\PradRatio = 1$. The red color represents the optimization problem for $E_{12} = 0$, and the blue curve is the problem with envelope correlation coefficient $E_{12} \leq 0.25$.}\label{pic:Cellphone4PortTotEffAndECC}
\end{figure}

Figure~\ref{pic:Cellphone4PortTotEffAndECC} shows the solution to problem~\eqref{eq:OptTask4Cellphone} where the set of red curves corresponds to $E_\T{max} = 0$ and the blue curves represent a solution for $E_\T{max} = 0.25$. The cost of zero envelope correlation coefficient~$E_{12}$ is a significant drop in total efficiency~$\toteff$ in the entire frequency region. Allowing a higher envelope correlation coefficient than zero (always the case in practical applications) leads to substantially better results in total efficiency.

\begin{figure}[t]
\centering
\includegraphics[width=0.48\textwidth]{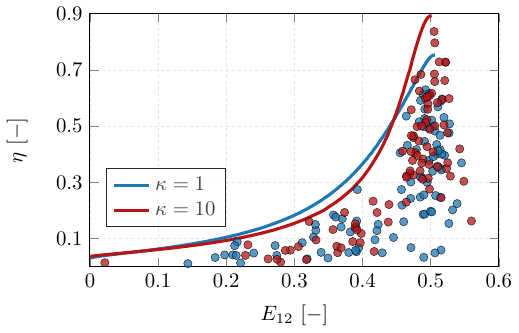}
\caption{Trade-off between the envelope correlation coefficient and total efficiency for the mobile terminal model at frequency $f=960\,$MHz. Two different ratios~$\PradRatio$ between powers are shown. The blue and red dots are randomly generated solutions from the Monte Carlo method. Pareto-optimal curves are highlighted by the solid lines.}\label{pic:Cellphone4Port2ClusterTradeOffTotEffECC}
\end{figure}

The trade-off between total efficiency~$\toteff$ and envelope correlation coefficient~$E_{12}$ is shown in Fig.~\ref{pic:Cellphone4Port2ClusterTradeOffTotEffECC} for~$f=960$\,MHz and for two different power ratios~$\PradRatio$. The maximum allowed value of correlation~$E_\T{max}$ is not limited in this case. The continuous lines are Pareto frontiers, while the dots represent randomly generated solutions. This example highlights the conflict between the power ratio and total efficiency caused by the lack of symmetries. Unlike the previous example, this antenna model has a significant trade-off between total efficiency and the envelope correlation coefficient.

\begin{figure}[t]
\centering
\includegraphics[width=0.48\textwidth]{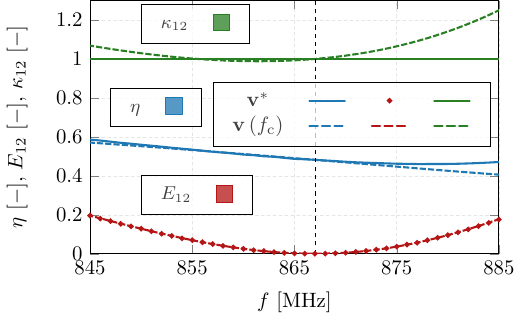}
\caption{Comparison of solutions to problem~\eqref{eq:OptTask4Cellphone} with $E_\T{max} = 0.25$. Solid lines show the performance of voltage vector $\pVvec$ optimized at each frequency. Dashed lines correspond to the voltage vector that is only optimal at frequency $f_\mathrm{c} = 867\,$MHz (highlighted by the black dashed vertical line).}\label{pic:Cellphone4PortSingleVector}
\end{figure}

To enable the full potential of antenna cluster technology, the distributed transceiver~\cite{art:Saleem2020IntegratedRFTransmitter} would be needed to set the optimal feeding coefficients at every frequency. Such an operation is impossible to implement due to physical restrictions~\cite{art:Hannula2018PerformanceAnalysisFreqReconfigAntCluster}. Therefore, the desired frequency band has to be divided into narrower sub-bands. The procedure~\eqref{eq:OptTask4Cellphone} is then applied for each sub-band only once. This opens several questions about the number of discretized sub-bands and the bandwidth of the optimal feeding vector.
Addressing this problem leads to further examination of optimal feeding coefficients near a single frequency. Figure~\ref{pic:Cellphone4PortSingleVector} shows the difference in performance for the case when the optimal solution to problem~\eqref{eq:OptTask4Cellphone} is applied to all frequencies in a given sub-band and for the case where the optimal solution for $f_\mathrm{c} = 867\,$MHz is applied to all considered frequencies. Total efficiency~$\toteff$, envelope correlation coefficient~$E_{12}$, and power ratio $\PradRatio = P_{22}/P_{11}$ at $f_\mathrm{c}$ are equal for both excitation schemes. All metrics gradually change as the distance from~$f_\mathrm{c}$ increases. To summarize, the optimal voltages do not deviate significantly when the frequency is changed, though the effect of frequency detuning must be evaluated in more detail in the future, also keeping in mind that feeding transmission lines between the feeding circuitry and the antenna port would result in additional dispersion. 

%%%%%%%%%%%%%%%%%%%%%%%%%%%%%%%%%%%%%%%%%%%%%%%%%%%%%%%%%%
%                       ########   
%                             #  
%                            #   
%                          ###   
%                          #   
%                         #   
%                        # 
%%%%%%%%%%%%%%%%%%%%%%%%%%%%%%%%%%%%%%%%%%%%%%%%%%%%%%%%%%
\section{Conclusion}
\label{sec:Conclusion}

The theory of antenna cluster optimization has been expanded toward evaluating the trade-off between total efficiency, envelope correlation coefficients, and power radiated by each cluster. The trade-off is evaluated via a solution to the quadratic optimization problem for maximization of total efficiency extended with the capability to deal simultaneously with the three mentioned parameters. This is enabled by introducing ratios between the power radiated by the clusters and radiated power component separation. The series of results in the two examples validate the proposed theory, with the second example highlighting the trade-off between the three mentioned parameters. The cellphone model confirms that the ground plane is the main radiator at lower frequencies. Total efficiency can be significantly increased when no requirement on a low envelope correlation coefficient is imposed. The most significant outcome of this paper is to demonstrate the ability of an antenna cluster to balance the level of far-field pattern correlation and total efficiency. 

The optimization procedure yields optimal voltages which provide Pareto-optimal performance when connected to the antenna's ports. Applying optimal voltages to all frequencies leads to an unattainable complexity of feeding circuitry. Therefore, the distributed transceiver must operate in sub-bands where one solution is applied. The feeding circuitry is also limited to discrete levels of feeding coefficients magnitudes and phases, which can further affect the performance, and with dispersion caused by transmission lines between the distributed transceiver and antenna ports. The outlined problem opens a question of the bandwidth of the optimal feeding vector and whether it should be added as an additional constraint to the optimization problem. Further research has to be conducted in the future. This work did not assume any form of matching circuits for further total efficiency enhancement. This is an additional set of degrees of freedom to be considered by an antenna designer.

An unresolved issue is the occasional appearance of a duality gap when the dual formulation is employed to solve the optimization problem. The employment of different solving schemes should also be investigated to guarantee a globally optimal solution. This problem will likely be enhanced when more than two antenna clusters are studied. Lastly, the presented investigation raised the question concerning the minimal number of ports required to guarantee the desired performance in all optimized metrics.

\appendices
%%%%%%%%%%%%%%%%%%%%%%%%%%%%%%%%%%%%%%%%%%%%%%%%%%%%%%%%%%
%                            #    
%                           # #   
%                          #   #  
%                         #     # 
%                         ####### 
%                         #     # 
%                         #     # 
%%%%%%%%%%%%%%%%%%%%%%%%%%%%%%%%%%%%%%%%%%%%%%%%%%%%%%%%%%
\section{Matrix Operators and Port Quantities}
\label{sec:MatrixOperators}

The considered radiating system is described by the impedance matrix $\Zmat = \RmatFS + \I\XmatFS$ extracted from any integral equations~\cite{book:VolakisIntEqua,book:ChewIntEqua} method 
\begin{equation}
\Zmat\Ivec = \Vvec,
\end{equation}
where~$\Ivec$ aggregates current approximation coefficients and $\Vvec$ is a discrete form of an excitation~\cite{book:HarringtonFieldComp}. The antenna metrics are commonly expressed as linear and quadratic forms of current vector $\Ivec$. Notable examples used in this paper are time-averaged radiated~$P$ and reactive~$\Preact$ powers
\begin{equation}
\frac{1}{2} \IvecH\Zmat\Ivec \approx P + \I\Preact.
\label{eq:AppAPrad}
\end{equation}
This paper deals with only lossless structures. However, the losses can be involved by adding loss matrix~$\RmatL$ into the impedance matrix $\Zmat$ without any other changes.

Two operators are defined for the purposes of transforming into port modes~\cite{art:Capek2020FindOptTotalRefCoef} with the first reading
\begin{equation}
D_{mn}= 
  \begin{cases}
    \displaystyle\, \zeta_{m} & m = n, \\
    \displaystyle\, 0 & \text{otherwise},
  \end{cases}
\end{equation}
where $\zeta_{m}$ is a parameter used to control units of port quantities. The second operator is a port indexing matrix defined as
\begin{equation}
N_{mn} =
  \begin{cases}
    \displaystyle\, 1 & \text{the $n$-th port is placed at the $m$-th position}, \\
    \displaystyle\, 0 & \text{otherwise}.
  \end{cases}
\end{equation}
Current vector~$\Ivec$ can now be controlled with voltage sources at antenna ports as
\begin{equation}
\Ivec = \Ymat\Dmat\Pmat\pVvec, \label{eq:ItoMultv}
\end{equation}
where admittance matrix~$\Ymat$ is an inverse of the~$\Zmat$ operator. Relation~\eqref{eq:ItoMultv} can be used to transform MoM matrices~$\M{M}$ to their port equivalents
\begin{equation}
\M{m} = \Pmat^\herm\Dmat^\herm\Ymat^\herm\M{M}\Ymat\Dmat\Pmat.\label{eq:OperatorTransform}
\end{equation}
As an example, relation~\eqref{eq:OperatorTransform} can be used to transform radiation operator~$\RmatFS$ from~\eqref{eq:AppAPrad} to port radiation matrix
\begin{equation}
\pGmatFS = \Pmat^\herm\Dmat^\herm\Ymat^\herm\RmatFS\Ymat\Dmat\Pmat,
\end{equation}
with the help of which the radiated power can be evaluated as
\begin{equation}
P \approx \frac{1}{2} \pVvec^\herm \pGmatFS \pVvec.
\label{eq:AppAPradPort}
\end{equation}

Another essential quantity is the power delivered into the radiating system which is used to constrain the optimization problem. If the signal generators are uncoupled and connected to the antennas with transmission lines of real-valued impedance, incident power can be determined as
\begin{equation}
\Pin = \frac{1}{2}\pAvecH\pAvec.\label{eq:Pin}
\end{equation}
To make it suitable for the optimization considered in this paper, incoming power waves~$\pAvec$ are related to port voltages~$\pVvec$ as~\cite{art:Capek2020FindOptTotalRefCoef}
\begin{equation}
\pAvec = \pKimat\pVvec = \frac{1}{2}\left(\Lambda^{-1} + \Lambda\pYmatFS\right)\pVvec. \label{eq:AvecToVvec}
\end{equation}

Finally, a combination of~\eqref{eq:AppAPradPort} and~\eqref{eq:Pin} leads to total efficiency~\cite{art:Capek2020FindOptTotalRefCoef}
\begin{equation}
\toteff = \frac{\IvecH\RmatFS\Ivec}{\pAvecH\pAvec} = \frac{\pVvecH\pGmatFS\pVvec}{\pVvecH\pKimat^\herm\pKimat\pVvec},
\end{equation}
which is the metric to be maximized in this text.

%%%%%%%%%%%%%%%%%%%%%%%%%%%%%%%%%%%%%%%%%%%%%%%%%%%%%%%%%%
%                         ######  
%                         #     # 
%                         #     # 
%                         ######  
%                         #     # 
%                         #     # 
%                         ######  
%%%%%%%%%%%%%%%%%%%%%%%%%%%%%%%%%%%%%%%%%%%%%%%%%%%%%%%%%%
\section{Power Waves Formulation}
\label{sec:PowerWavesForm}
The theory introduced above is general and can be adapted to arbitrary types of port quantities. This section shows how to use it with incident power waves~$\pAvec$. An embedded radiation matrix~\cite{art:Stein1962OnCrossCoupling} reads
\begin{equation}
\Gamma_{mn} = \frac{1}{4\pi a_m^*a_n}\oint\limits_{\varOmega}\V{F}_{\T{e},m}^*\left(\theta,\varphi\right)\V{F}_{\T{e},n}\left(\theta,\varphi\right)\T{d}\varOmega,
\end{equation}
where $\V{F}_{\T{e},m},\V{F}_{\T{e},n}$ represent embedded far-field patterns obtained with exciting $m,n$-th port with incoming power wave $a_m,a_n$ and terminating rest of the ports into matched load. If the system has no Ohmic losses, then it can be obtained from the relation
\begin{equation}
\M{\Gamma} = \M{E} - \pSmat^\herm\pSmat,
\end{equation}
where $\M{E}$ is the identity matrix of corresponding size, and matrix $\pSmat$ contains s-parameters of the radiating system.

Radiated power, in this case, is calculated as
\begin{equation}
\Prad = \frac{1}{2}\pAvecH\M{\Gamma}\pAvec.
\end{equation}
Total efficiency reads
\begin{equation}
\toteff = \frac{\Prad}{\Pin} = \frac{\pAvecH\M{\Gamma}\pAvec}{\pAvecH\pAvec},
\end{equation}
and embedded far-field pattern correlation reads
\begin{equation}
\rho_{\T{e},mn} =\frac{\pAvecH_m\M{\Gamma}_{mn}\pAvec_n}{\sqrt{\pAvecH_m\M{\Gamma}_{mm}\pAvec_m}\sqrt{\pAvecH_n\M{\Gamma}_{nn}\pAvec_n}}.\label{eq:FarFieldCorrelationPowerWaves}
\end{equation}
Both quantities can be improved either by changes in the radiating body or by feeding optimization, for which can be used the routine defined above. The only difference is, that terms $\Prad_{mn}$ are defined as
\begin{equation}
P_{mn} = \frac{1}{2}\pAvecH_m\M{\Gamma}_{mn}\pAvec_n.
\end{equation}
It is worth mentioning that while the total efficiency is equal for power waves and corresponding voltages~\eqref{eq:AvecToVvec}, the envelope correlation coefficients~\eqref{eq:FarFieldCorrelationPowerWaves} and~\eqref{eq:FarFieldCorrelationVolt} are different quantities because of distinct excitation.

Port radiation matrix~$\pGmatFS$ and embedded radiation matrix~$\M{\Gamma}$ are related as\begin{equation}
\M{\Gamma} = \pKimat^{-\herm}\pGmatFS\pKimat^{-1}.
\end{equation}
Matrix~$\M{\Gamma}$ can be extracted from an arbitrary field solver making the whole framework generalized and independent of the chosen method.

%%%%%%%%%%%%%%%%%%%%%%%%%%%%%%%%%%%%%%%%%%%%%%%%%%%%%%%%%%
%                          #####  
%                         #     # 
%                         #       
%                         #       
%                         #       
%                         #     # 
%                          #####  
%%%%%%%%%%%%%%%%%%%%%%%%%%%%%%%%%%%%%%%%%%%%%%%%%%%%%%%%%%
\section{Far-field Correlation}
\label{sec:FarFieldCorr}

Mutual radiated power~$P_{mn}$, typically used to quantify far-field orthogonality, can be evaluated as
\begin{equation}
P_{mn} = \frac{1}{2Z_0}\oint\limits_{S}\V{F}^*_m\sphArg\cdot\V{F}_n\sphArg\D\varOmega,
\end{equation}
where indices~$m,n$ denote current vectors~$\M{I}_m,\M{I}_n$ excited on an antenna. Utilizing method of moments together with~\eqref{eq:AppAPradPort}, this can be written as a quadratic form
\begin{equation}
P_{mn} = \frac{1}{2}\IvecH_m\RmatFS\Ivec_n = \frac{1}{2}\pVvecH_m\pGmatFSind{m}{n}\pVvec_n,
\end{equation}
with the help of which a far-field correlation coefficient~$\rho_{mn}$ is expressed as 
\begin{equation}
\rho_{mn} = \frac{\pVvecH_m\pGmatFSind{m}{n}\pVvec_n}{\sqrt{\pVvecH_m\pGmatFSind{m}{m}\pVvec_m}\sqrt{\pVvecH_n\pGmatFSind{n}{n}\pVvec_n}}.\label{eq:FarFieldCorrelationVolt}
\end{equation}

Although the generally complex coefficient~\eqref{eq:FarFieldCorrelationVolt} is not directly suitable for convex optimization, individual quadratic forms in the nominator and denominator can be added to the optimization problem as additional constraints.

The envelope correlation coefficient $E_{mn}$ reads
\begin{equation}
\ECC_{mn} \approx \left\vert \rho_{mn}\right\vert^2.
\end{equation}

%%%%%%%%%%%%%%%%%%%%%%%%%%%%%%%%%%%%%%%%%%%%%%%%%%%%%%%%%%
%                         ######  
%                         #     # 
%                         #     # 
%                         #     # 
%                         #     # 
%                         #     # 
%                         ######  
%%%%%%%%%%%%%%%%%%%%%%%%%%%%%%%%%%%%%%%%%%%%%%%%%%%%%%%%%%
\section{Correlation Expressed with Power Ratios}
\label{sec:SystemAnalysis}
This section expresses correlation through power ratios. Considering a case of two antenna clusters $M=2$, the far-field correlation coefficient reads
\begin{equation}
\rho_{12} = \frac{\pVvecH_1\pGmatFSind{1}{2}\pVvec_2}{\sqrt{\pVvecH_1\pGmatFSind{1}{1}\pVvec_1}\sqrt{\pVvecH_2\pGmatFSind{2}{2}\pVvec_2}} = \frac{1}{2}\frac{\beta_{12}+\I\gamma_{12}}{\sqrt{\alpha_{11}}\sqrt{\alpha_{22}}}\label{eq:FarFieldCorrelationPowerCoef},
\end{equation}
which is obtained by dividing every term $\pVvecH_m\pGmatFSind{m}{n}\pVvec_n$ with $\pVvecH\pGmatFS\pVvec$ and utilizing the power ratios. The ratio between self powers $\PradRatio = P_{11}/P_{22}$ is usually of interest. Using~\eqref{eq:PowerCoefSummation} and ratio~$\PradRatio$ allows us to express $\alpha_{11}$ as 
\begin{equation}
\alpha_{11} = \frac{1 - \beta_{12}}{1 + \PradRatio},\label{eq:AlphaExpressBeta}
\end{equation}
and $\alpha_{22} = \PradRatio\alpha_{11}$. Putting~\eqref{eq:AlphaExpressBeta} into relation~\eqref{eq:FarFieldCorrelationPowerCoef} leads to
\begin{equation}
E_{12} = \left(\frac{1 + \PradRatio}{2\sqrt{\PradRatio}}\right)^2\frac{\beta_{12}^2+\gamma_{12}^2}{\left(\beta_{12}-1\right)^2}.\label{eq:EnvelopeCorrelationApprox}
\end{equation}
The first term in parentheses in~\eqref{eq:EnvelopeCorrelationApprox} is the scaling factor and the rest of the expression forms the equation for the ellipse. Considering $\PradRatio = 1$ for simplicity, then the expression~\eqref{eq:EnvelopeCorrelationApprox} can be rewritten as
\begin{equation}
\frac{\left(1 - E_{12}\right)^2}{E_{12}} \left(\beta_{12} + \frac{E_{12}}{1-E_{12}}\right)^2 + \frac{1 - E_{12}}{E_{12}}\gamma_{12}^2 = 1, 
\end{equation}
which is an ellipse equation. The general $\PradRatio$ would need only a re-scaling of $E_{12}$ by the factor in parentheses in equation~\eqref{eq:EnvelopeCorrelationApprox}.

\begin{figure}[t]
\includegraphics[scale=1]{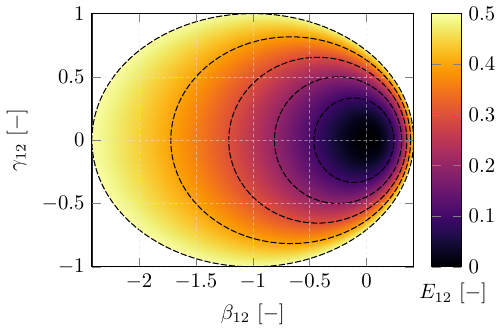}
\caption{Depiction of envelope correlation expressed with power ratios for the case of two antenna clusters and $\PradRatio=1$. Dashed ellipses show curves of the constant envelope correlation coefficient. Values higher than $E_{12} > 0.5$ are not shown in compliance with~\cite{web:3GPP}.}\label{pic:RelativePowerBetaGamma}
\end{figure}

%%%%%%%%%%%%%%%%%%%%%%%%%%%%%%%%%%%%%%%%%%%%%%%%%%%%%%%%%%
%                         #######  
%                         #      
%                         #      
%                         #####  
%                         #      
%                         #      
%                         #######  
%%%%%%%%%%%%%%%%%%%%%%%%%%%%%%%%%%%%%%%%%%%%%%%%%%%%%%%%%%
\section{Power Ratios Intervals}
\label{sec:RadPowerRatios}
For a simplification of the matrix description, the indexing operators $\Cmat{m}\in\mathbb{B}^{N\times N_m}$ are introduced as
\begin{equation}
C_{m,kl} = 
\begin{cases}
    \displaystyle 1 & \text{the $k$-th port is the $m$-th cluster $l$-th port}, \\
    \displaystyle 0 & \text{otherwise}.
  \end{cases}
\end{equation}
Applying this operator from both sides on the port-mode radiation matrix gives
\begin{equation}
\pGmatFSind{m}{n} = \Cmat{m}^\herm\pGmatFS\Cmat{n}.
\end{equation}
Using the combination of $\Cmat{m}\Cmat{m}^\herm$ on the same group of operators provides a matrix that reads
\begin{equation}
\widetilde{\M{g}}_{0,mn} = \Cmat{m}\Cmat{m}^\herm\pGmatFS\Cmat{n}\Cmat{n}^\herm =
\begin{bmatrix}
\M{0} & \hdots & \M{0} \\
\vdots & \pGmatFSind{m}{n} & \vdots \\
\M{0} & \hdots & \M{0}
\end{bmatrix}.\label{eq:CmatCmatOperator}
\end{equation}

Terms~\eqref{eq:RadPowerTerms} can be expressed as 
\begin{equation}
P_{mn} = \frac{1}{2}\pVvec^\herm\Cmat{m}\Cmat{m}^\herm\pGmatFS\Cmat{n}\Cmat{n}^\herm\pVvec,
\end{equation}
which can be then directly utilized in the implementation of the problem~\eqref{eq:OptTaskConstrained}.

Relation \eqref{eq:CmatCmatOperator} can be used for establishing feasible intervals for power ratios by means of generalized eigenvalue problems. Starting with self powers, the maximal value of $\alpha_{mm}$ is found as 
\begin{equation}
\widetilde{\M{g}}_{0,mm} \pVvec_j = \alpha_{j}\pGmatFS\pVvec_j,
\end{equation}
and the corresponding interval reads $\alpha \in \left[0,\max{\alpha_{j}}\right]$. 

A similar approach is used for mutual power ratios. The generalized eigenvalue problems read
\begin{align}
\left(\widetilde{\M{g}}_{0,mn} + \widetilde{\M{g}}_{0,nm}\right)\pVvec_k &= \beta_{k}\pGmatFS\pVvec_k, \label{eq:BetaBounds} \\
\I\left(\widetilde{\M{g}}_{0,mn} - \widetilde{\M{g}}_{0,nm}\right) \pVvec_l &= \gamma_{l}\pGmatFS\pVvec_l, \label{eq:GammaBounds}
\end{align}
and determine intervals $\beta_{mn} \in \left[\min{\beta_{k}},\max{\beta_{k}}
\right]$ for the real part and $\gamma_{mn} \in \left[\min{\gamma_{l}},\max{\gamma_{l}}
\right]$ for the imaginary part of mutual powers.


\end{document}